# Analytical Loss Factors in Approximation of the Leontovich Boundary Conditions.

S.S. Baturin[1], A.D. Kanareykin[1,2]

[1]. St.Petersburg Electrotechnical University LETI, St.Petersburg Russia 197376

[2]. Euclid Techlabs LLC, Solon, OH 44139

**Abstract:**

Recently the new method of the Cherenkov fields and loss factors of a point-like electron bunch passing through longitudinally homogeneous structures lined with arbitrary slowdown layers was proposed [1]. It was shown that the Cherenkov loss factor of the short bunch does not depend on the waveguide system material and is a constant for any given transverse dimensions and cross-section shapes of the waveguides. It was shown that with the proposed approach one can use a relatively simple method for the calculation of the total loss factor using an integral relation based on the cylindrical slowdown waveguide model. With this paper, we demonstrate that the same integral relation that we call "relativistic Gauss theorem" can be applied in case impedance boundary conditions (IBC) also known as Leontovich boundary conditions.

**Introduction.**

Relativistic, high intensity and small emittance electron bunches are the basis of linear collider (ILC [2], CLIC [3]) and FEL (LCLS, X-FEL etc.) [4] projects among many others. These bunches excite Cherenkov wakefields as long as electrons pass through the accelerating structures or other longitudinally extended components of a beam line (pipes, collimators, bellows). Theoretical analysis of Cherekov radiation commonly considers a "short bunch" approach [5-8]. It can be applied for various Cherenkov generation parameters, where the moving charge size is much less that the fundamental wavelength if the high frequency spectrum range is not under investigation.

Recently a new theoretical approach that can be used for obtaining direct analytical formulas for electromagnetic field components at the position of a point-like bunch was proposed [1]. It was shown that the loss factor of the point-like bunch does not depend on the waveguide system material and is a constant for any given transverse dimensions and cross-section of the waveguides. The equivalence and exact matching of the loss factor of beams passing through various waveguide configurations was analyzed too. The proposed approach considers using an integral relation based on the cylindrical slow wave structure model. For planar, square and other cross-section geometries one can obtain the loss factor by using a conformal mapping from the solution for the cylindrical case [1].

At the same time, for many practical applications an impedance boundary conditions (IBC) or Leontovich conditions [9-12] are commonly used. Recently it was proposed [13] to verify if the proposed method can be applied for impedance boundary conditions. With this paper, we have shown that the Leontovich boundary conditions can be also used if the relativistic Gauss theorem is applied.

**Integral transformation and Leontovich boundary conditions.**

Let us consider the circulation of the magnetic field on the metal boundary of a waveguide and apply Green's theorem

$$\oint \mathbf{H} \cdot d\mathbf{l} = \iint_{S_\perp} (\nabla \times \mathbf{H}) \cdot d\mathbf{S}, \qquad (1)$$

where the integral on the right hand side is calculated over the cross section $S_\perp$ of the waveguide, and the left hand side integral is taken along the metal sleeve of the waveguide. Let's write the integral of the right hand side of (1) in scalar form, using Ampère's law:

$$\nabla \times \mathbf{H} = \frac{\partial \mathbf{D}}{\partial t} + \rho \mathbf{V}; \qquad \iint_{S_\perp} (\nabla \times \mathbf{H}) \cdot d\mathbf{S} = \iint_{S_\perp} \left( \frac{\partial D_z}{\partial t} + \rho V \right) dS. \qquad (2)$$

Here $\rho = q\delta(z-Vt)\delta(x-x_0)\delta(y-y_0)$ is the charge density $\mathbf{V} \uparrow\uparrow e_z$ - charge velocity vector.

Let us consider a special case when slow down media could be replaced with the Leontovich conditions [12] on the boundary of the vacuum channel. We consider a circulation of magnetic field on the boundary of the vacuum channel and rewrite it with the help of Green's theorem (1) and Ampère's law (2) as:

$$\varepsilon_0 \iint_{S'_\perp} \frac{\partial E_z^V}{\partial \zeta} dS = -q\delta(\zeta) + \frac{1}{c} \oint_{l'} \mathbf{H} \cdot d\mathbf{l} \qquad (3)$$

Here $S'_\perp$ - is the cross section of the vacuum channel, $l'$ - is the vacuum gap cross section boundary, $\zeta = Vt - z$. Leontovich conditions could be written as [12]

$$\varepsilon_0 c \sqrt{\varepsilon} \mathbf{E} = \sqrt{\mu} \mathbf{H} \times \mathbf{n} \qquad (4)$$

Here $\mathbf{n}$ - is the normal vector to the boundary. In case of cross section orthogonal to the beam direction $H_\tau$ ($\mathbf{H}$ on contour $l'$) from (4) with external normal vector could be expressed as

$$H_\tau^V = -\varepsilon_0 c \sqrt{\frac{\varepsilon}{\mu}} E_z^V, \qquad (5)$$

where upper index $V$ corresponds to vacuum channel. Substitution of (5) into (3) leads to

$$\iint_{S'_\perp} \frac{\partial E_z^V}{\partial \zeta} dS + \sqrt{\frac{\varepsilon}{\mu}} \oint_{l'} E_z^V dl = -\frac{q}{\varepsilon_0} \delta(\zeta) \qquad (6)$$

Integrating (6) over $\zeta$ and taking into account condition at infinity $E_z^V(-\infty) = 0$, one can write

$$\iint_{S'_\perp} E_z^V(\zeta) dS + \sqrt{\frac{\varepsilon}{\mu}} \oint_{l'} \int_{-\infty}^{\zeta} E_z^V(\zeta') d\zeta' dl = -\frac{q}{\varepsilon_0} \theta(\zeta) \qquad (7)$$

In the limit $\gamma \to \infty$ $E_z$ field in front of the bunch $\zeta < 0$ should be equal to zero, thus second integral in (7) could be rewritten as:

$$\lim_{\gamma \to \infty} \sqrt{\frac{\varepsilon}{\mu}} \oint_{l'} \int_{-\infty}^{\zeta} E_z^V(\zeta') d\zeta' dl = \sqrt{\frac{\varepsilon}{\mu}} \oint_{l'} \int_{0}^{\zeta} E_z^V(\zeta') d\zeta' dl \qquad (8)$$

Now we take limit of $\zeta \to 0$, in this case one can rewrite (7) with (8) as

$$\iint_{S'_\perp} E_z^V(0+\Delta\zeta)dS + \sqrt{\frac{\varepsilon}{\mu}}\oint_{l'} E_z^V(0)\Delta\zeta\, dl = -\frac{q}{\varepsilon_0}\theta(0+\Delta\zeta) \tag{9}$$

here $\Delta\zeta \to 0$. Proceeding with the limit one can write:

$$\iint_{S'_\perp} E_z^V(0)dS = -\frac{q}{2\varepsilon_0} \tag{10}$$

$$\iint_{S'_\perp} E_z^V(0^+)dS = -\frac{q}{\varepsilon_0} \tag{11}$$

As one can see Leontovich boundary conditions in considered limits gives quite the same results (10), (11) as general approach or integral relation [1].

**Frequency Dispersive Media.**

In case of a dispersive medium Leontovich conditions in Fourier space could be written as [12]

$$\varepsilon_0 c\sqrt{\varepsilon(\omega)}\tilde{\mathbf{E}} = \sqrt{\mu(\omega)}\tilde{\mathbf{H}}\times\mathbf{n} \tag{12}$$

In case of cross section orthogonal to the beam direction $H_\tau$ ($\mathbf{H}$ on contour $l'$) from (12) with external normal vector could be expressed as:

$$\tilde{H}_\tau^V = -\varepsilon_0 c\sqrt{\frac{\varepsilon(\omega)}{\mu(\omega)}}\tilde{E}_z^V \tag{13}$$

Here $\tilde{\mathbf{E}}$ and $\tilde{\mathbf{H}}$ is the Fourier components of the electric and magnetic field vectors respectively.

$$\tilde{\mathbf{E}}(\omega) = \frac{1}{2\pi}\int_{-\infty}^{\infty} \mathbf{E}\exp(i\omega t)dt \tag{14}$$

In Fourier space equation (3) with (13) could be written as:

$$-i\omega\iint_{S'_\perp}\tilde{E}_z^V dS + \sqrt{\frac{\varepsilon(\omega)}{\mu(\omega)}}\oint_{l'}\tilde{E}_z^V dl = -\frac{q}{2\pi\varepsilon_0}\exp\left(i\omega\frac{z}{V}\right) \tag{15}$$

Inverse Fourier transform of (15) leads to:

$$\iint_{S'_\perp}\frac{\partial E_z^V}{\partial t}dS + \oint_{l'}\int_{-\infty}^{\infty} g(t-t')E_z^V(Vt'-z)dt'dl = -\frac{q}{\varepsilon_0}\delta(Vt-z) \tag{16}$$

Here we have denoted:

$$g(t) = \int_{-\infty}^{\infty}\sqrt{\frac{\varepsilon(\omega)}{\mu(\omega)}}\exp(-i\omega t)d\omega \tag{17}$$

At the point of a charge $z = Vt$. Substitution of $z = Vt$ into (16) leads to

$$\iint_{S'_\perp} \frac{\partial E_z^V}{\partial t} dS + \oint_{l'} \int_{-\infty}^{\infty} g(t-t') E_z^V (Vt'-Vt) dt' dl = -\frac{q}{\varepsilon_0} \delta(0) \qquad (18)$$

In the limit $\gamma \to \infty$ $E_z$ field in front of the bunch should be equal to zero $E_z(Vt-z) = 0$, $Vt-z < 0$. This leads to $E_z(Vt'-Vt) = 0$, $Vt'-Vt < 0 \Rightarrow E_z = 0$, $t' < t$. Using this fact one may rewrite (18) as:

$$\iint_{S'_\perp} \frac{\partial E_z^V}{\partial t}\bigg|_{z=Vt} dS + \oint_{l'} \int_{t}^{\infty} g(t-t') E_z^V (Vt'-Vt) dt' dl = -\frac{q}{\varepsilon_0} \delta(0) \qquad (19)$$

Now if we assume that charge is passing the channel for a long time enough or, in other words we are interested in wake asymptotic solution only then one should proceed with the limit of $t \to \infty$. This leads to:

$$\lim_{t \to \infty} \oint_{l'} \int_{t}^{\infty} g(t-t') E_z^V (Vt'-Vt) dt' dl \to 0 \qquad (20)$$

Thus for the point of the charge in the limit $t \to \infty$ one can write:

$$\iint_{S'_\perp} \frac{\partial E_z^V}{\partial t}\bigg|_{z=Vt} dS = -\frac{q}{\varepsilon_0} \delta(0) \qquad (21)$$

Let us introduce new coordinate $\zeta = Vt - z$ and rewrite (21) in terms of $\zeta$:

$$\iint_{S'_\perp} \frac{\partial E_z^V}{\partial \zeta}\bigg|_{\zeta=0} dS = -\frac{q}{\varepsilon_0} \delta(0) \qquad (22)$$

As a consequence of the fact that field in front is zero and from (22) one can achieve:

$$\iint_{S'_\perp} E_z^V (0) dS = -\frac{q}{2\varepsilon_0} \qquad (23)$$

$$\iint_{S'_\perp} E_z^V (0^+) dS = -\frac{q}{\varepsilon_0} \qquad (24)$$

**Wakefield calculation using the Leontovich conditions and the integral theorem.**

In this section we will consider a direct solution of the wakefield analytical simulations with Leontovich boundary conditions in case of the point-like charge passing along the axis of the cylindrical vacuum channel lined with dielectric [14]. We compare here the previously obtained solution with those that could be derived with the help of the Leontovich boundary conditions, formula (6) by applying recently proposed integral theorem. Following the steps of [14] one can derive the following formula for the Fourier image of the scalar potential:

$$\Phi(\omega, r) = \frac{q}{4\pi^2 \varepsilon_0 V} [K_0(\chi r) + \alpha I_0(\chi r)] \qquad (25)$$

Here $\chi = \omega/V\sqrt{1-\beta^2}$ and $\alpha$ is given by:

$$\alpha = \frac{\sqrt{\mu}\beta K_1(\chi a) + i\sqrt{\varepsilon(1-\beta^2)}K_0(\chi a)}{\sqrt{\mu}\beta I_1(\chi a) - i\sqrt{\varepsilon(1-\beta^2)}I_0(\chi a)} \tag{26}$$

Longitudinal electric field $E_z$ could be found as [14]

$$E_z(\omega, r) = \frac{i\omega(1-\beta^2)}{c}\Phi(\omega, r) \tag{27}$$

In the limit $\gamma \to \infty$ and $\mu = 1$ $E_z(\omega, r)$ could be written as:

$$E_z(\omega, r) = \frac{q}{4\pi^2 \varepsilon_0 c} \frac{2i}{a^2 k - 2ia\sqrt{\varepsilon}} \tag{28}$$

Inverse Fourier transform leads to

$$E_z(\zeta, r) = \frac{q}{2\pi^2 \varepsilon_0 a^2} \int_{-\infty}^{\infty} \frac{i\exp(i\zeta k)dk}{k - 2i\sqrt{\varepsilon}/a} \tag{29}$$

Here $\zeta = Vt - z$ and $k = \omega/c$. Integral (29) has one pole at $k = 2i\sqrt{\varepsilon}/a$ thus field could be written as:

$$E_z(\zeta, r) = -\frac{q}{\pi\varepsilon_0 a^2}\theta(\zeta)\exp\left(-\frac{2\sqrt{\varepsilon}}{a}\zeta\right) \tag{30}$$

One can from (30) see that

$$E_z(0, r) = -\frac{q}{2\pi\varepsilon_0 a^2}, \tag{31}$$

which corresponds to a general result [1]. Now let's consider equation(6), where taking into account the fact that in case $\gamma \to \infty$ in cylindrical vacuum channel with the charge moving along axis z then $E_z$ does not depend on transverse coordinates, and we immediately have from (6):

$$\frac{\partial E_z^V}{\partial \zeta}\iint_{S'_\perp} dS + E_z^V \sqrt{\frac{\varepsilon}{\mu}} \oint_{l'} dl = -\frac{q}{\varepsilon_0}\delta(\zeta), \tag{32}$$

which leads to

$$\frac{\partial E_z^V}{\partial \zeta} + E_z^V \frac{2}{a}\sqrt{\frac{\varepsilon}{\mu}} = -\frac{q}{\pi a^2 \varepsilon_0}\delta(\zeta) \tag{33}$$

In case $\mu = 1$ solution of this equation could be written as

$$E_z(\zeta,r) = -\frac{q}{\pi\varepsilon_0 a^2}\theta(\zeta)\exp\left(-\frac{2\sqrt{\varepsilon}}{a}\zeta\right) \tag{34}$$

One can see that (34) exactly matches previous result (30) [14].

Correspondingly, it was shown that for the waveguide with a slowdown layer the solutions for the wakefields obtained as a Fourier decomposition or by applying an integral theorem [1] are identical even if the Leontovich approximation (4) is used.

**Summary**

In conclusion, with this paper we demonstrated that the impedance boundary conditions (IBC) or Leontovich conditions can be applied along with the recently proposed integral "relativistic Gauss theorem" for the point-like charge. First of all, it was shown that proposed theorem can be derived directly from the Maxwell equations even if only Leontovich approximation is used instead of the standard boundary conditions. The same approach was extended to the solutions for the dispersive media as well. Finally, the wakefield simulations for the Leontovich conditions were carried out for the slowdown waveguide using a Fourier decomposition method and proposed integral theorem. Both methods concluded with the same well-known expression for the wakefield behind the bunch.

**References:**

1. S.S.Baturin, A.D. Kanareykin. Cherenkov Loss Factor of Short Relativistic Bunches: General Approach. http://arxiv.org/abs/1308.6228
2. ILC Technical Design Report, http://linearcollider.org.
3. M. Aicheler. A Multi-TeV Linear Collider Based on CLIC Technology: CLIC Conceptual Design Report, http://cds.cern.ch/record/1500095?ln=en
4. C Bostedt et al. J. Phys. B 46, 164003, 2013.
5. K.F.Bane. Wakefields of sub-picosecond electron bunches. SLAC-pub-11829, (2006).
6. K.L.F. Bane, G. Stupakov. Nucl. Instr. Meth. Phys. Res. A, 690, p.106, (2012).
7. A. Novokhatski, A. Mosnier, Proc.of PAC 1997, Vancouver, 1997, p. 1661, (1997)
8. K. Bane and G. Stupakov, Phys. Rev. ST-Accel. Beams 6, 024401 (2003).
9. A. Chao. In book: Physics of Collective Beam Instabilities in High Energy Accelerators, Wiley & Sons, New York (1993).
10. B.W.Zotter and S.A.Kheifets. Impedances and Wakes in High-Energy Accelerators. (World Scientific,. Singapore, 1997).
11. S.V. Yuferev, N. Ida. Surface Impedance Boundary Conditions: A Comprehensive Approach. CRC-Press, 2010.
12. L. D. Landau and E. Lifshitz, Electrodynamics of Continuous Media, Pergamon, London, 1960
13. G.Stupakov. Private communications.
14. M. Bolotovskii, Usp. Fiz. Nauk, 75, p. 295 (1961); Sov.Phys. Usp, 4, p. 781 (1962).